 \documentclass[preprint2]{aastex}

\newcommand{\scs}{\scriptsize}
\newcommand{\til}{$\sim$}

\newcommand{\ergsec}{\thinspace\hbox{$\hbox{erg}\thinspace\hbox{s}^{-1}$}}
\def\spose#1{\hbox to 0pt{#1\hss}}
\def\simlt{\mathrel{\spose{\lower 3pt\hbox{$\mathchar"218$}}
     \raise 2.0pt\hbox{$\mathchar"13C$}}}
\def\simgt{\mathrel{\spose{\lower 3pt\hbox{$\mathchar"218$}}
     \raise 2.0pt\hbox{$\mathchar"13E$}}}

\newcommand{\decsec}[2]{$#1\mbox{$''\mskip-7.6mu.\,$}#2$}

\newcommand{\decsectim}[2]{$#1\mbox{$^{\rm s}\mskip-6.3mu.\,$}#2$}

\newcommand{\aminbyamin}[2]{#1$'\times$#2$'$}

\def\today{\ifcase\month\or
January\or February\or March\or April\or May\or June\or
July\or August\or September\or October\or November\or December\fi
\space\number\day, \number\year}

\slugcomment{Revised version submitted to the Astronomical Journal \today}

\shorttitle{The Optical Identification of the X-ray Burster in NGC\thinspace6441}
\shortauthors{Homer et al.}

\begin{document}


\title{The Optical Identification of the X-ray Burster X1746-370 in the Globular Cluster NGC\thinspace6441 \footnote{\ Based on observations with
the NASA/ESA
Hubble Space Telescope, obtained at the Space Telescope Science Institute,
which is operated by the Association of Universities for Research in
Astronomy, Inc., under NASA contract NAS5-26555.}
}


\author{L. Homer and Scott F. Anderson}
\affil{Astronomy Department, Box 351580, University of Washington, Seattle, WA 98195-1580}
\email{homer@astro.washington.edu; anderson@astro.washington.edu}
\author{Bruce Margon and Ronald A. Downes}
\affil{Space Telescope Science Institute, 3700 San Martin Drive, Baltimore, MD 21218}
\email{margon@stsci.edu ; downes@stsci.edu}
\and
\author{Eric W. Deutsch}
\affil{Institute for Systems Biology, 1441 N 34th St, Seattle, WA 98103-8904}
\email{edeutsch@systemsbiology.org}




\begin{abstract}
We find convincing observational evidence to confirm the optical identification of the X-ray burster X1746-370 located in the globular cluster
NGC\thinspace6441. {\it Chandra}/HRC-I imaging yields a much improved X-ray position for the source, which we show to be fully
consistent with our rederived position of a UV-excess star, U1, in the same astrometric reference frame.  In addition, the smaller {\it Chandra} X-ray error circle excludes the only other blue stars previously identified in the old {\it Einstein} circle.  We have also obtained {\it Hubble Space
Telescope}/STIS time-resolved optical spectra of star U1.  Although there are no strong line features, the flux distribution demonstrates U1 to
be unusually bright in the blue and faint in the red, consistent with earlier WFPC2 photometry.   More notably, the flux level of the continuum is seen to vary significantly compared to
stars of similar brightness.  Indeed, the lightcurve can plausibly be fit by a 5.73 hr period sinusoid, which is the period of the recurring
X-ray dips seen in this source. The presence of modulations in both wavelengths strengthens the case for an orbital origin, and therefore deepens the
puzzle of the 
unusual energy independent X-ray dips. Lastly, we note that X1746-370 remains the longest period confirmed X-ray burster in a globular cluster,
and
the only one 
with a period typical of the galactic population as a whole.

\end{abstract}

\keywords{globular clusters: individual (NGC\thinspace6441) --- stars: neutron ---
X-rays: bursts --- X-rays: stars}

\section{INTRODUCTION}
\label{sect:intro}
Globular clusters are expected to provide ideal environments for the formation of close binaries, with their high stellar densities and much enhanced
rates of star interaction.   This is certainly the case for the X-ray bright interacting systems.  Almost from the advent of
X-ray astronomy it has been known that the cluster population of luminous ($\simgt10^{36}$\ergsec) low-mass X-ray binaries (LMXBs) is $\simgt$100 times enhanced (per unit
stellar mass) relative to the galaxy as a whole  \citep{clar75,katz75}.  Another curious fact is that 11 of these 13 LMXBs must contain neutron stars rather
than black holes, as they exhibit type-I X-ray bursts (understood as thermonuclear runaway burning on the compact object's surface).
Moreover, the sensitive X-ray imaging of the {\it Chandra X-ray Observatory} has now revealed an equal number of probable
quiescent systems in the few clusters examined to date \citep{grin01a,grin01b,hein01,home01d,pool02,rutl02}, as well as the existence of {\em two} persistent LMXBs in M15 \citep{whit01}.

	The study of globular cluster LMXBs benefits greatly from a multi-wavelength approach.  X-ray data, in general, only probe the vicinity of
the central source, apart from the highest inclination systems where  material farther
out can cross the line of sight.  The periods of four cluster LMXBs have been determined from eclipses by the donor star and/or from the periodic dips in their X-ray flux,
understood as due to obscuration by vertically extended material near the edge of the accretion disc.  However, with the identification of
optical/UV counterparts (in all but one case requiring the resolution of {\it HST}), we can immediately begin to estimate the linear scale of a system from the 
$L_X/L_{opt}$ ratio, which has been shown to scale with disc area \citep{vP94}.  Photometric monitoring has also proven effective in revealing
variability on the binary period, whilst the optical spectra can in principle provide definitive corroboration of a counterpart
and further useful diagnostics.

The X-ray burster  X1746-370, located in NGC\thinspace6441, is one of the X-ray ``dippers''.  From a continuous {\it
EXOSAT} observation, \citet{parm89} first observed dips and inferred an orbital
periodicity, which was refined to 5.73$\pm0.15$hr  by \citet{sans93} using a more extensive {\it Ginga} dataset.  Apart from a single deep
(90\% flux decrease) dip, which showed spectral hardening \citep{jonk00}, all the observed dips have been shallow (\til15\%) and have shown no clear energy
dependence. The apparently energy independent dipping is puzzling, implying that the obscuring material responsible for the electron
scattering has metal abundance $\simlt 0.01$ times solar, but this seems unlikely given the close to solar metallicity of the cluster as a
whole \citep{djor93}. The alternative explanations are: photoionization of the material, a number of varying spectral components conspiring
together, or an extended X-ray source (i.e. an accretion disc corona).  However, the most recent broad-band spectroscopy of
\citet{parm99} using {\it BeppoSAX} argues against any of these possibilities.  One might contend that the standard dipping interpretation itself could be
erroneous-- and it is certainly true that none of the X-ray period determinations has been sufficiently precise to confirm the recurrent period as
orbital in origin based on its stability. Additional progress has been made in the optical.  Using {\it HST}/WFPC2 imaging data, \citet{deut98} identified
a variable, UV-excess star (designated U1) in the {\it Einstein} X-ray error circle.  However, given the surprisingly large number of similar UV-bright stars in
the cluster, there remained a possibility that U1 might be a chance superposition on the X-ray position. The {\it a posteriori} probability of
this coincidence was calculated to be
\til30\% (based upon a 3\arcsec\ radius 90\% confidence {\it Einstein} error circle).

As part of a continuing program to probe the optical/UV counterparts to the luminous globular cluster X-ray sources, we have reexamined the
optical position in the light of new {\it Chandra} X-ray imaging data, and also obtained
time-resolved {\it
HST}/STIS optical spectra of the candidate counterpart to the burster X1746-370. 

\section{DATA ANALYSIS AND RESULTS}

\subsection{{\it Chandra} X-ray Position}
{\it Chandra} observed the field of X1746-370 with a short exposure of 3.2ks on 2000 May 5; we obtained the dataset from the archive.  The \til\aminbyamin{30}{30} field-of-view high resolution
camera imager \citep[HRC-I;][]{murr97} was approximately centered on the cluster center. Data reduction was undertaken with routines in {\tt
CIAO}\footnote{Available at http://asc.harvard.edu}. The observation was not affected by any periods of background flaring, hence applying the
standard good time intervals and correcting for dead time, we obtained 2903s of data on source. Owing to this shallow exposure, only three sources were detected at greater than 3$\sigma$ above background in the
entire field, using a Mexican-Hat wavelet source detection routine ({\tt wavdetect}). The brightest source (50000 counts) was clearly the main target, whilst
the other two were very much fainter (only 27 and 11 net counts) {\em and} were located within 3\arcsec\ and 20\arcsec\ of chip edges
respectively, where the PSF is most extended and distorted.  As a consequence, neither faint source could be reliably used to improve/confirm the nominal astrometry from the satellite's aspect solution, although probable
optical identifications (within \til1\arcsec) to the bright (non-cluster) stars HD 161892 and GSC{\scs II} S2220122126 were made. For the reprocessed version of the dataset we used, there are no known aspect offsets,
hence we take the centroided position of X1746-370 given by {\tt wavdetect} as the best X-ray position possible: $\alpha=17^{\rm h}50^{\rm m}$\decsectim{12}{73}$\pm$\decsectim{0}{05}, $\delta=-37^{\circ}03'$\decsec{06}{8}$\pm$\decsec{0}{6}, where we 
quote our uncertainties as those of the {\it Chandra} aspect \citep{aldc00}.

	We have estimated our sensitivity limits for other lower luminosity cluster sources. Within the cluster half-mass radius of 38\arcsec\
\citep{harr96}, but beyond the \til20\arcsec\ wings of X1746-370, a source with $L_X\sim10^{33}$\ergsec would be just detectable, yielding $\simgt$5
counts\footnote{ We assume a distance of 11.2 kpc and
$N_H=2.1\times10^{21}$cm$^{-2}$ \citep[as determined from {\it BeppoSAX} spectra,][]{parm99}, and typical blackbody or thermal
bremsstrahlung spectral models representing quiescent LMXB or cataclysmic variable star emission.}, comparable to the brightest quiescent LMXBs
detected by {\it Chandra} in other clusters (see \S\ref{sect:intro}). Closer to the bright source the limits will naturally be higher due to
the increase in 
effective background; this includes the entirety of the cluster core region.

\subsection{Refined {\it HST} Optical Position}
Although a precise optical position for the proposed counterpart to X1746-370 (star U1) was published by \citet{deut98}, that astrometry was
based on the
{\it HST Guide Star Catalog} reference frame \citep{lask90}.  This predates the improvement made possible by {\it Hipparcos}, and the construction of the International
Celestial Reference System \citep[ICRS;][]{hog00}, the reference frame for the {\it Chandra} astrometry.  Although the difference in a given position is typically $\simlt1$\arcsec, this is still significant
when compared to the sub-arcsecond precision of the {\it Chandra} X-ray position.  Hence we have redetermined the astrometry  using the very accurate USNO-A2.0 star
catalog \citep{mone98}, which makes it
possible to tie an arbitrary field rather easily to the ICRS with sub-arcsecond precision. Once again we make use of the ground-based CCD image of
the NGC\thinspace6441 field kindly provided by G. Jacoby \citep[e.g. see][]{jaco97} . We select 46 bright, isolated stars in common between
the USNO-A2.0 catalog (epoch 1982.0 in this field) and the ground-based CCD image
(epoch 1995.6) and fit an astrometric solution to the image using IDL
procedures written by E.W.D. and from the {\it Astronomy User's Library}
\citep{Land93}.  The residuals of the fit ($\sigma$=\decsec{0}{56})
imply an approximate uncertainty ($\sigma/\sqrt{n-3}$) in the alignment
to the USNO-A2.0 frame of \decsec{0}{09} before considering proper
motion effects. \citet{deut99} derived empirical uncertainties in the transfer of the USNO A-2 frame (via reference star matching) to 
ground-based images comprising the Second Digitized Sky Survey (DSS-II), which effectively
includes the scatter induced by the differing epochs of the data and random proper motion effects.  Hence, we conservatively adopt the 1$\sigma$ radial
uncertainty of \decsec{0}{35} given in the last row of
Table 1 in \citet{deut99}.

The next step is to transfer this solution to the U-band (F336W filter) images from the {\it HST}/WFPC2 (epoch 1994.7).  These data are
described in detail
in \citet{deut98}. The PC was centered approximately on the core, and hence the corresponding portion of the ground-based image is very severely
crowded. It is therefore clearly advantageous  to use
the data from the WF chips outside of the cluster core and relate this to PC chip on which the candidate counterpart star U1 is located.  This
is made possible by the STSDAS routine {\tt metric}, which uses the well-calibrated relative positions of each of the chips and the
geometrical distortions across the field of each to accurately give an RA and Dec for any star on any chip, based upon the astrometric solution present in the WF2 image header.  We therefore identified 52 well-isolated stars which appear on both the
ground-based image and the 3 WF chips.  Given the ICRS astrometry solution now written into the header of the ground-based image, we are able to
obtain ICRS positions for each of these stars and cross-compare to the nominal result given by {\tt metric} for the WFPC2 astrometric
solution.  We find that shifts of \decsec{1}{6} and \decsec{0}{001} in RA and Dec are adequate to bring the WFPC2 data onto the ICRS, with no correction to the orientation
necessary.  Again from the standard deviation of the residuals we estimate a radial uncertainty of \decsec{0}{02} for this step, negligible compared to
the other uncertainties. With the corrections in place, we measure a new position for star U1 in the ICRS of:
$\alpha=17^h50^m$\decsectim{12}{728}$\pm$\decsectim{0}{029}, $\delta=-37^{\circ}03'$\decsec{06}{53}$\pm$\decsec{0}{35}. This position differs
appreciably ($>1$\arcsec) from that quoted by \citet{deut98}, with essentially all the difference due to the different astrometric frames, and
should supercede those earlier data.

\subsection{Optical Spectroscopy}
On 1999 June 28, we obtained {\it HST} optical spectroscopy of star U1.  We used STIS \citep{wood98} with a \decsec{0}{2}$\times$52\arcsec\ aperture which was
well-centered on the target, following a blind  offset from a nearby bright star.  The G430L
grating was employed, together with the STIS-CCD detector, yielding a useful wavelength range of \til3000\AA--5700\AA\ and spectral
resolution  $\approx5$\AA. Five {\it HST} orbits of
data were taken, with five separate exposures during each, except for the first {\it HST} orbit for which there are only four exposures. 
\begin{figure}[!htb]
\resizebox{.45\textwidth}{!}{\rotatebox{-90}{\plotone{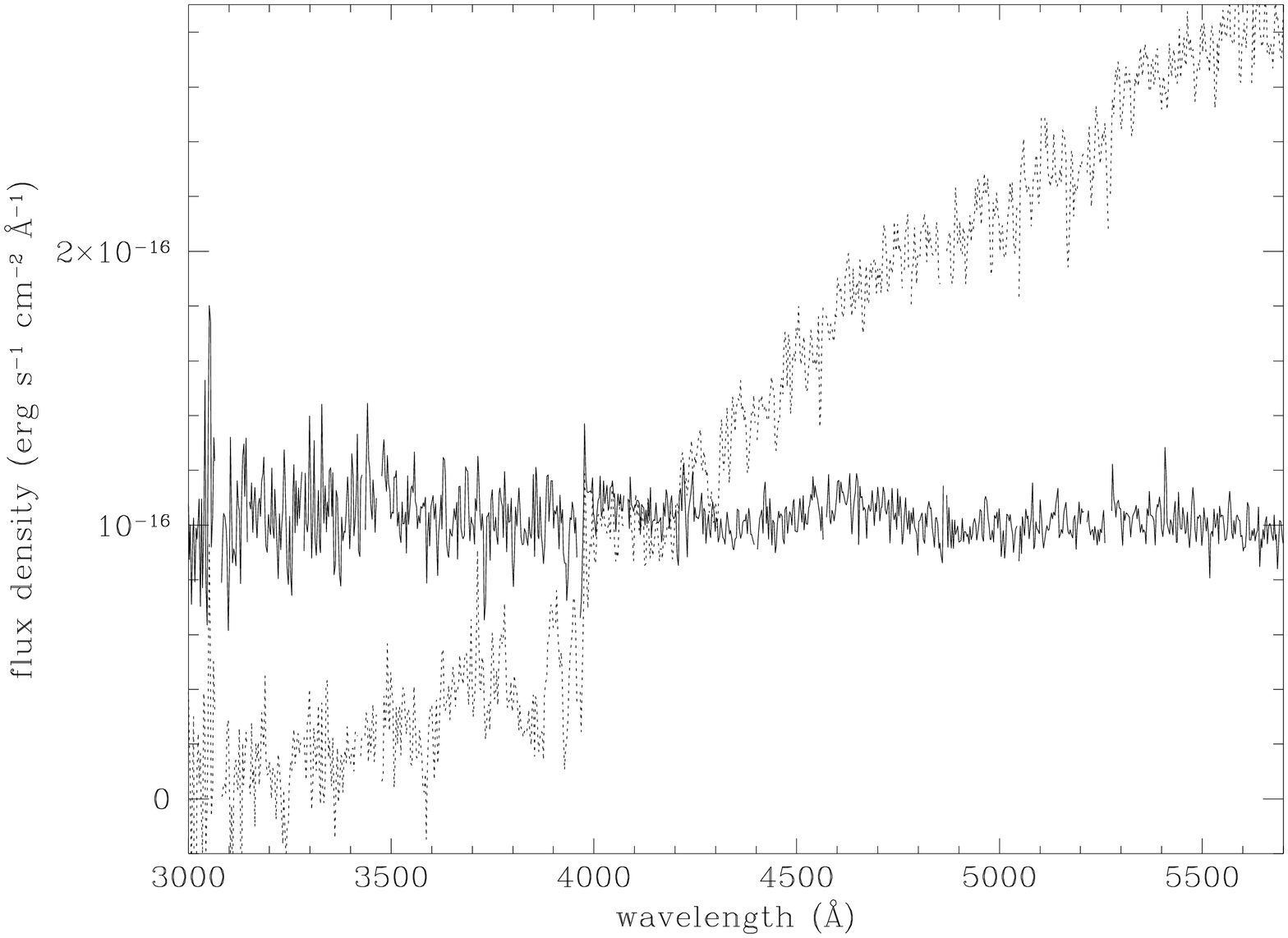}}}
\resizebox{.45\textwidth}{!}{\rotatebox{-90}{\plotone{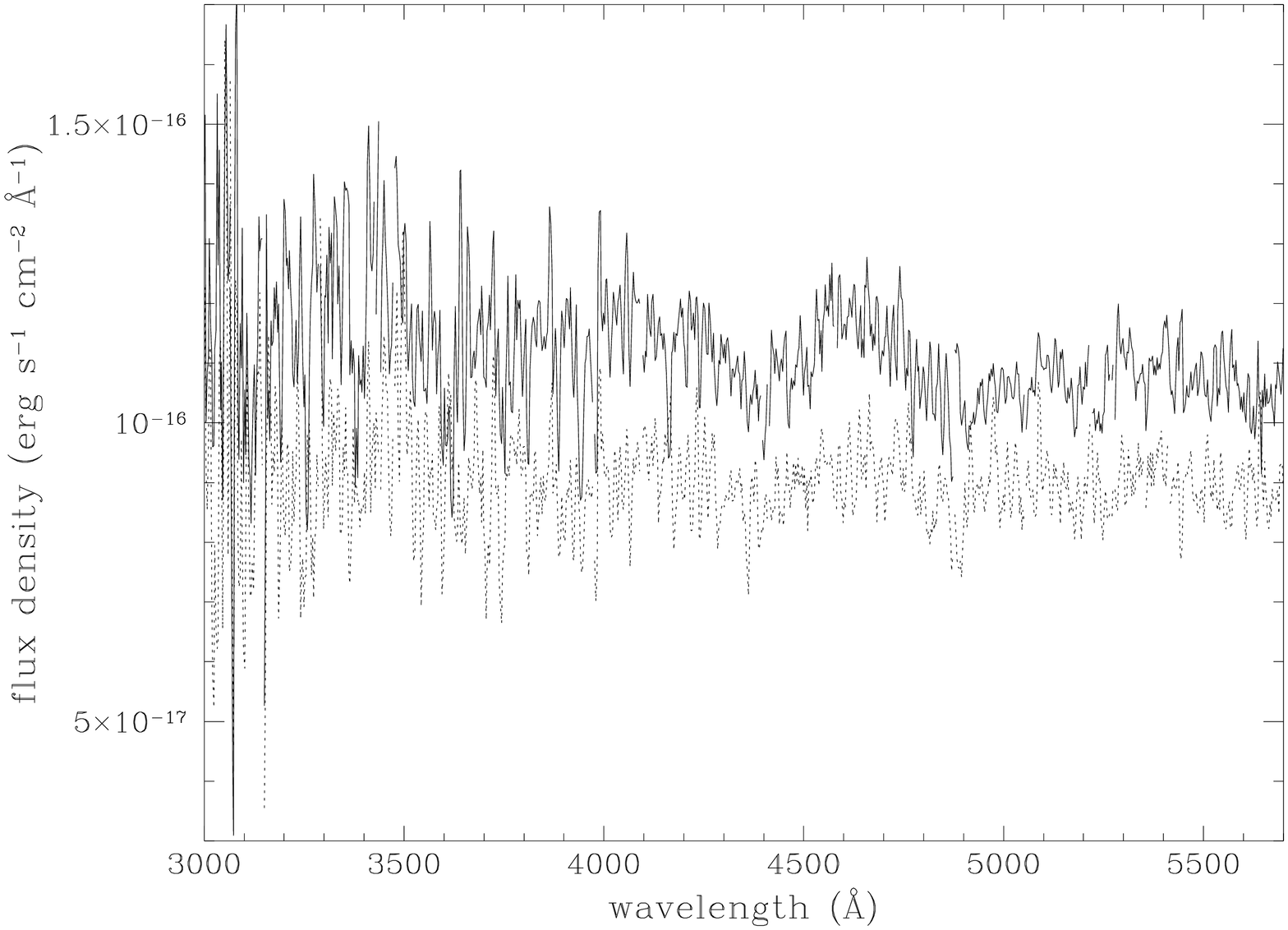}}}
\caption{{\it HST}/STIS optical spectra (not dereddened).  Upper: from all five {\it HST} orbits of data -- the candidate counterpart to the X-ray
burster, star U1 (solid), with a (typically red) comparison star 
over-plotted (dashed).  With approximately constant flux density, U1 is relatively extremely blue.  Lower: spectra of U1 from two
distinct orbits when it was at its 
brightest and faintest( to aid clarity the data have been boxcar smoothed over 3 pixels), demonstrating the change in flux levels, visible even in these dispersed spectra.  We also note the lack of
appreciable changes in spectral slope, implying that we have obtained 
reasonably contamination free spectra.  Note: the gaps in the spectra are regions where the data were
corrupted by CCD defects.\label{fig:spec}}
\end{figure}

The standard pipeline reductions produce a variety of data products, where all the data taken in a given {\it HST} orbit have been combined together.  This combining
facilitates the removal of the numerous cosmic ray events, via one-sided sigma clipping (together with a noise model for the detector).
Unfortunately, the CCD also suffers from a number of defects, which appear as anomalously high or low spikes in the extracted spectra.  These
we identified and later removed by hand from the spectra. Taking the combined cosmic-ray rejected spectrum images from each {\it HST}
orbit, we applied the STSDAS {\tt x1d} aperture extraction and calibration routine to produce 1-D flux and wavelength calibrated 
spectra for the target and a number of other reasonably isolated stars of similar brightness located on the long slit.  For each star, we kept the width of the extraction regions
the same for all images, but allowed the position to track any change in the trace position.  We were careful in 
setting both the source and neighboring background regions to exclude as much as possible contributions from other cluster stars, but inevitably
in this dense core region, many very faint stars will be unavoidable.  Examination of these spectra shows that: (i) U1 possesses a featureless
continuum with no strong emission lines or notable absorption features (apart from Ca{\scs II} H+K from imperfect background subtraction) at the modest signal-to-noise of these data; (ii) in comparison to other stars the (reddened) continuum of U1
has almost constant flux density, making it the brightest star in the blue ($\simlt4000$\AA) but one of the faintest in the red; (iii) relative
to the comparison stars the flux level of the continuum of U1 appears to vary
significantly. In figure~\ref{fig:spec} 
we show the median stacked spectra (total exposure time of 12680s) of star U1 and a representative comparison star, and two individual spectra
showing the variability of U1 between its highest and lowest
flux levels. As this cluster is substantially reddened ($E_{B-V}\sim0.4$), the observed flat spectrum implies intrinsic colors that are very
blue. 

In order to compare the flux levels to the previous WFPC2 photometry, we have estimated the corresponding STMAG magnitudes from our spectra: for F336W
($U$) they range from  $18.72\pm0.02$ to $18.94\pm0.05$  
and for F439W ($B$) from $18.76\pm0.02$ to $18.91\pm0.05$, with uncertainties based on systematic variations (see below). Comparing to Table 1 of
\citet{deut98}, the F336W magnitudes are consistent with the 1994 levels given the source's variability, but at F439W the source is \til0.3 mag
brighter than in 1994, i.e. it appears considerably redder.  It is possible that there is spectral contamination present from a redder star (e.g. the
wings of a 
neighbor, centered just off the slit, \decsec{0}{35} away), but the lack
of significant color-terms in the variability suggests not (see fig.\ref{fig:spec} lower panel).  Moreover, we note that the 1995 WFPC2 data found the source with similar F439W brightness
levels to our 1999 spectrophotometry, and since no F336W data were taken in 1995, the source could have been similarly redder then. 

In order to further quantify the variability exhibited by U1, we have calculated the total net counts in the continua of U1 and several
comparison stars 
for each {\it HST} orbit.  We repeated the aperture extractions from the combined cosmic-ray rejected images, but this time without applying any
further calibrations, yielding count rate versus pixel spectra.  Next we fitted the continuum shape of the
source and background spectra with spline functions, and replaced
points deviant by more than 4$\sigma$, by the appropriate continuum fit. Lastly, we summed the counts in these cleaned 1-D spectra over all
pixels. At this stage we also estimated the random errors in the net counts, based upon Poisson
counting statistics together with CCD read-noise contributions.  These indicated that a high nominal accuracy can be obtained, and therefore that variations seen
in the comparison stars (which should of course exhibit constant fluxes) must arise from systematic effects. 
\begin{figure}[!htb]
\resizebox{.45\textwidth}{!}{\rotatebox{-90}{\plotone{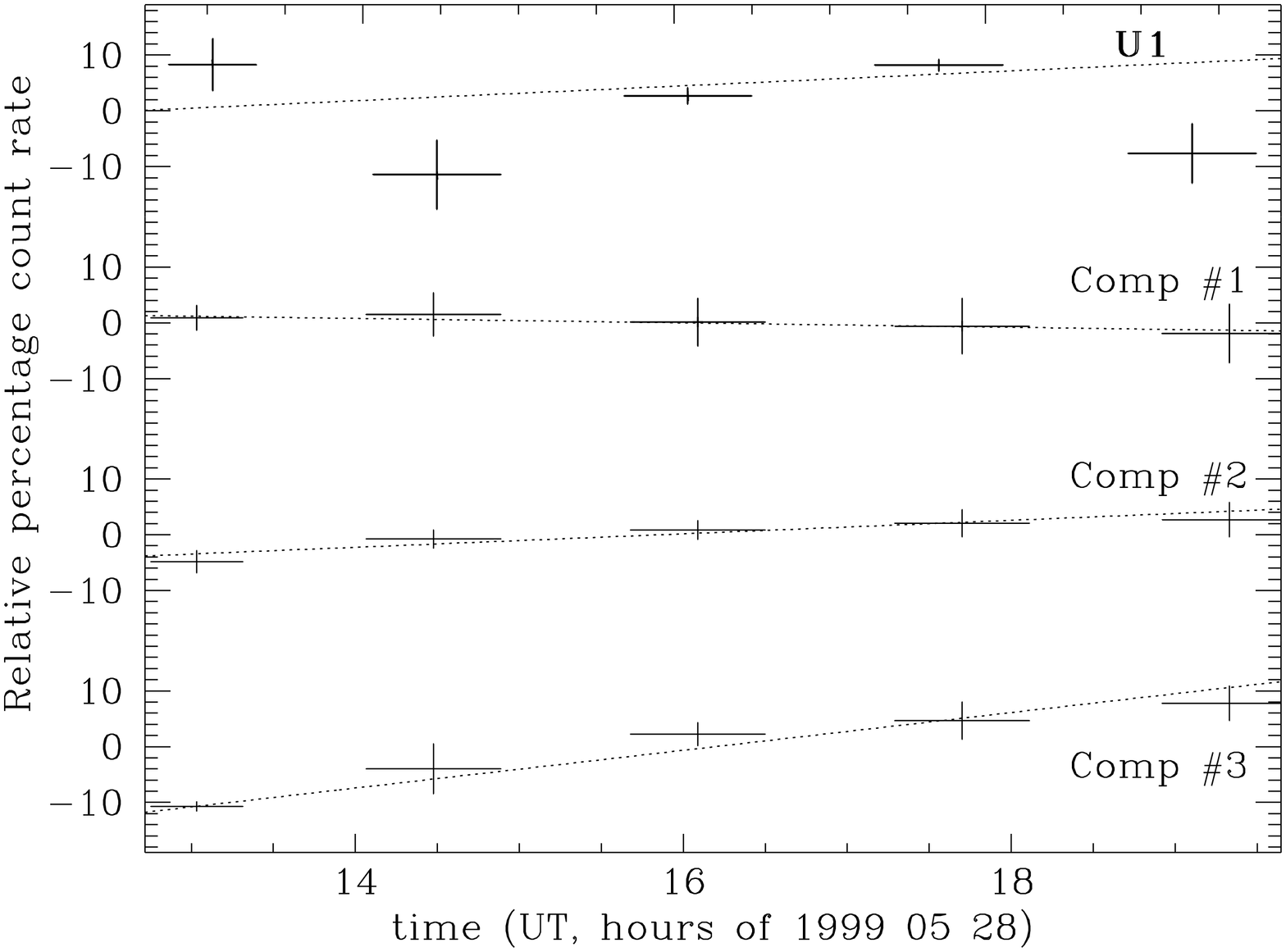}}}
\resizebox{.45\textwidth}{!}{\rotatebox{-90}{\plotone{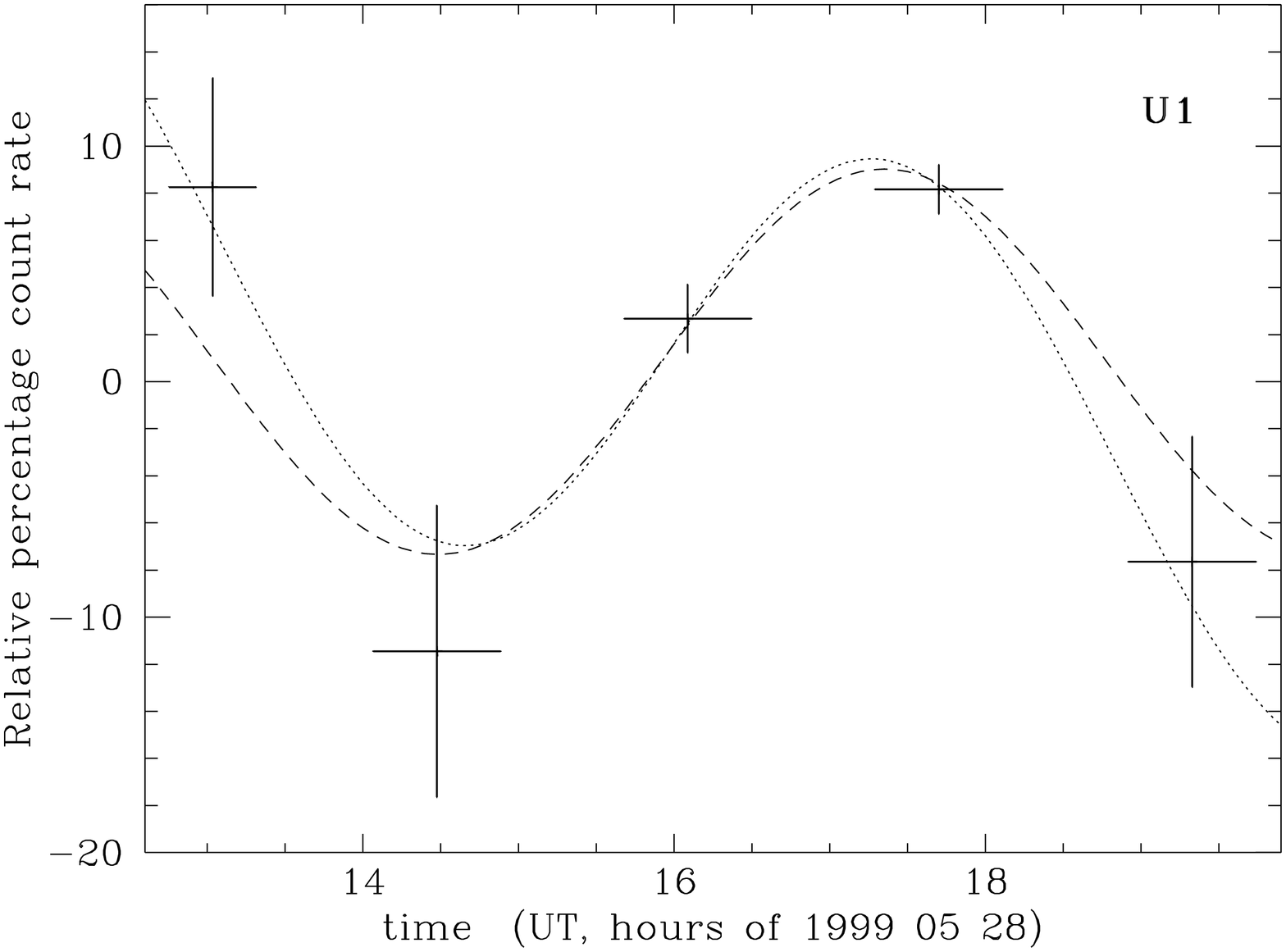}}}
\caption{Lightcurves for the candidate counterpart star U1 and three other stars of similar brightness; datapoints are derived from the orbitally
stacked spectrum images and (systematic) errors estimated from higher time resolution analysis.  Upper: a linear trend has been fitted to each curve,
but is notably a very poor fit to the data for star U1.  Lower: considering star U1 only, two models have been fitted: (i) a constant + sinusoid (dashed)
with period constrained to 5.73 hr (X-ray dip period), (ii) the same, but with a linear term added (dotted).\label{fig:lc1}}
\end{figure}

To examine these systematic effects further we considered higher time-resolution data. Instead of combining all the exposures from a given {\it HST}
orbit to remove cosmic rays, we used all available adjacent pairs (i.e. calculating a running average) giving 3 or 4 cleaned spectra per
orbit.  Once again systematic trends were apparent.  At least two statistically independent data points are available per orbit, so we use the
scatter of the most divergent pair to conservatively estimate the uncertainty in the count rates we obtained from the
fully combined data.  The final
lightcurves are plotted for U1 and the comparison stars in figure~\ref{fig:lc1}. Systematic long-term trends in count rate are immediately
apparent, but these are reasonably well fit (i.e. $\chi_\nu^2<1$) by a linear function in every case apart from U1, where $\chi_\nu^2=7$. In the lower panel, we plot the data for U1 once more, and this time fit sinusoidal models to the data, one with a constant offset and
another allowing for a linear trend, akin to that of the comparison stars. In each case, the period is
constrained to that of the X-ray dips (5.73 hr), and we see that the observed optical modulation is indeed consistent with this periodic model. The fits are clearly much
better than for the simple linear
trend, with $\chi_\nu^2=1.8$ and 0.9 for the sinusoid, without and with a linear term respectively. Applying an F-test to the two models indicates that the addition of the linear term to the
sinusoid is only marginally preferred at the 66\% confidence level.  We find peak-peak amplitudes
of 16\% and 20\%, in reasonable agreement with the short term \til30\% changes seen previously in this star in the F336W photometry
\citep{deut98}. While we certainly cannot claim to have detected the X-ray period in the optical data, we feel our data do provide evidence for
this effect.

\begin{figure}[!htb]
\resizebox{.5\textwidth}{!}{\plotone{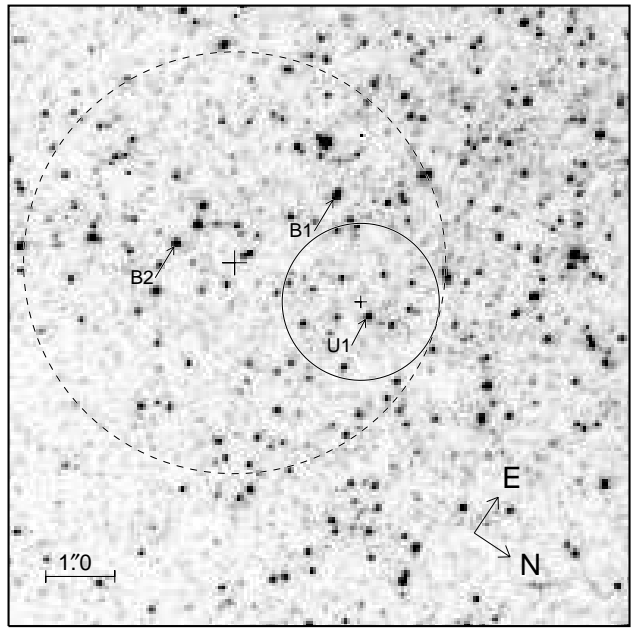}}
\caption{Comparison of the various positions for X1746-370 and its optical counterpart.  An F336W PC image from {\it HST} is shown, with the old {\it Einstein}
 error
circle (dashed) and new {\it Chandra} circle (solid) overlaid (both 90\% confidence).  The two blue stars identified in the larger error circle
by \citet{deut98} as well as
their proposed UV-bright counterpart, U1, are indicated.  We note the excellent positional agreement between the reduced X-ray error region
and the position of U1.\label{fig:posns}}
\end{figure}
\begin{table*}[!htb]
\caption{X-ray and optical positions for X1746-370 (in the ICRS).\label{tab:posns}}
\begin{tabular}{l l l l l l} 
\tableline\tableline
Band & Telescope &$\alpha$ (J2000) & Error & $\delta$ (J2000) & Error\\
&&(h, m, s)& ($1\sigma$)&($^\circ$, \arcmin, \arcsec)&($1\sigma$)\\
\tableline
X-ray & {\it Chandra}/HRC-I & 17 50 12.73 & \decsectim{0}{05} & -37 03 06.8 & \decsec{0}{6}\\
Optical &{\it HST}/WFPC2& 17 50 12.728& \decsectim{0}{029} & -37 03 06.53& \decsec{0}{35}\\
\tableline
\end{tabular}
\end{table*} 

\section{DISCUSSION}
The new {\it Chandra} X-ray and optical positions for X1746-370 and its proposed counterpart U1 are summarized in table~\ref{tab:posns}, and illustrated in figure~\ref{fig:posns}.  The positional
agreement is excellent and well-within the (small) estimated radial uncertainties, providing strong support for the optical
identification. Following the {\it a posteriori} probability estimate of \citep{deut98}, but with a new 90\% confidence radius of \decsec{1}{1} for the
{\it Chandra} error circle, the area enclosed is 7 times smaller than for {\it Einstein}, and hence only a $\sim$4\% probability remains that
we have chance alignment of an unrelated UV-excess cluster star with the X-ray position.  Moreover, our new {\it HST}/STIS spectra confirm that
star U1 is unusually blue. But more conclusively, the temporal coverage afforded by our
time-resolved spectra are sufficient to confirm the optical variability of the source, which appears periodic and can be well fitted with a 
sinusoid constrained to the 5.73 hr X-ray period.

The firm identification of the optical counterpart does have further implications for our view of the nature of the source, including the issue of the
unusual energy independent X-ray dips. First, the most likely origin for the sinusoidal optical modulation on the X-ray period is the varying contribution from
the bright X-ray heated face of the donor.  Hence, its possible detection would not only confirm the correct identification, but also support
the standard picture that the recurring dips are related to
periodic obscuration on the orbital period.  Second, as previously noted by  \citet{parm99}, the faintness of the optical star implies that
$L_X/L_{opt}\sim1000$, typical of LMXBs in which we directly observe the central source, and therefore consistent with the detection of bursts
from this source. In particular, this suggests that it is unlikely that the dips are primarily due to obscuration of an extended accretion disc corona (ADC); in
the classical ADC sources where only scattered X-rays are visible, $L_X/L_{opt}\sim20$.  Our optical spectral data also argue against such an
ADC interpretation.
Considering all LMXBs with comparable periods (and hence disc sizes), most show characteristic emission lines in this spectral region at
H$\beta$, He{\scs II} $\lambda$4686 and the Bowen C{\scs III}/N{\scs III} blend $\lambda$4640; moreover, all three known classical ADC sources exhibit
strong emission at one or more of these lines \citep[see e.g.][]{vP95}.  Hence, the lack of strong line emission in X1746-370 is unlike any of
these ADC systems, and is even unusual compared to the field LMXBs in general.  However, we note that our {\it HST} spectra of the globular cluster LMXBs in
NGC\thinspace6624, and 6712 \citep{deut98T} are similar to X1746-370 in NGC\thinspace6441 
insofar as they are also very blue/UV, but largely 
featureless at modest signal-to-noise and resolution.

Within the context of the diverse nature of the cluster LMXB population, X1746-370 might at first glance be considered a rather average system.  In terms of period
it lies midway between the three with ultra-short periods (P$<$ 1 hr) in NGC\thinspace6624, 1851 and 6712 \citep[see e.g.][and references
therein]{home01c}, and the two long period systems in Terzan 6 and M15 (AC211) \citep[P=12.4 hr and 17.1 hr
respectively;][]{intZ00,ilov93}. However, it is the longest period confirmed burster, since the Terzan 6 LMXB has not been seen to burst,
and we now know that AC211 (M15-X1) is almost certainly an ADC source and M15-X2, a source without a known orbital period, is the probable burster there \citep{whit01}. The recent
results on NGC\thinspace6652 \citep{hein01} indicate that its burster is perhaps the most similar, with a longest possible period of 4.4 hr, though 0.92
hr also fits the available data.  If this shorter period in NGC\thinspace6652 does turn out to be correct, of the five bursters with determined or well-constrained
periods, four would be double-degenerate ultra-compact systems and only X1746-370 would be similar to a typical galactic
burster. \citet{deut00} have already commented on this prevalence of very exotic systems in globular clusters.  It would appear that the unique formation/evolution processes at work in globular cluster cores \citep[see e.g.][for a review]{hut92}-- e.g. the tidal capture and exchange
encounter mechanisms and subsequent stellar interactions, leading to the hardening of already hard binaries -- may have led to an enhancement of the ultra-compact LMXBs at the expense of the wider
systems like X1746-370. In any case, the (growing) population of LMXBs may serve as important tracers of the stellar dynamics and evolution within globular clusters.

\acknowledgments
We once again thank George Jacoby and collaborators for providing ground-based CCD images of NGC\thinspace6441.  We are grateful to the
anonymous referee for useful comments. Support for this work was provided by NASA
through grant NAG5-7932, as well as ST\,ScI grant GO07363.01-96A, and SAO grant GO0-1011X.


\end{document}